\documentclass[a4paper]{article}

\usepackage{amssymb}
\usepackage{latexsym}

\usepackage{url}
\usepackage{xcolor}
\usepackage{graphicx}

\usepackage[margin=1.1in]{geometry}

\usepackage[style=numeric,backend=bibtex]{biblatex} 
\bibliography{pattern_bib}

\newcommand{\eg}{\emph{e.g.}, }       % for example
\newcommand{\ie}{\emph{i.e.}, }      % that is
\newtheorem{hypothesis}{Hypothesis}

\title{Limitations and Alternatives for the Evaluation of Large-scale Link Prediction}

\author{Garcia-Gasulla, D. \& Ayguad\'{e}, E. \& Labarta, J.\\
Barcelona Supercomputing Center (BSC)\\dario.garcia@bsc.es\vspace{8pt}\\
\& Cort\'{e}s, U.\\
Universitat Polit\`{e}cnica de Catalunya - BarcelonaTECH\vspace{8pt}}

\begin{document}
\maketitle

\begin{abstract}
{Link prediction, the problem of identifying missing links among a set of inter-related data entities, is a popular field of research due to its application to graph-like domains. Producing consistent evaluations of the performance of the many link prediction algorithms being proposed can be challenging due to variable graph properties, such as size and density. In this paper we first discuss traditional data mining solutions which are applicable to link prediction evaluation, arguing about their capacity for producing faithful and useful evaluations. We also introduce an innovative modification to a traditional evaluation methodology with the goal of adapting it to the problem of evaluating link prediction algorithms when applied to large graphs, by tackling the problem of class imbalance. We empirically evaluate the proposed methodology and, building on these findings, make a case for its importance on the evaluation of large-scale graph processing.}
\textit{Graph Mining \and Link Prediction \and Evaluation Methodology}
\end{abstract}

\section{Introduction}

The structural particularities of graphs (\ie networks) has motivated the design of specific machine learning methods for processing this type of data. These knowledge discovery tools typically try to exploit structural properties of high-dimensional, inter-connected data sets, with the goal of learning from the relational patterns of its entities. Among the names used to refer to some of these tools are:

\begin{itemize}
\item Graph-based data mining (\cite{GDM,GRPH})
\item Statistical Relational Learning (\cite{SRL,TENSOR})
\item Link Mining (\cite{LINK,LPCN})
\item Network or Link Analysis (\cite{GUI,COVDIR})
\item Network Science (\cite{NEWPERS})
\item Structural Mining (\cite{STRMIN})
\end{itemize}

For the sake of simplicity from now on we refer to all these methods using the general term \emph{graph mining}.

The characterization of graph mining algorithms is relevant, not only because graphs represent data in a rather unique way, but because they are also able to capture a different type of information. While traditional table representations of entity/value pairs naturally capture \emph{intra-entity patterns}, and so does traditional machine learning, network data captures mostly \emph{inter-entity} patterns. Mining graphs therefore requires a shift in perspective, moving from an instance-attribute paradigm to an instance-instance paradigm.

These new methods of machine learning were designed to tackle network related problems such as:

\begin{itemize}
\item Finding the relevance of entities based on their relations or \emph{link-based object ranking}\cite{LINK,lu2016vital} \eg PageRank (\cite{PR}) and HITS (\cite{HITS})
\item Finding groups of entities strongly related or \emph{community detection}\cite{fortunato2010community} \eg stochastic blockmodeling (\cite{STOCH})
\item Finding reoccurring association patterns or \emph{frequent subgraph discovery}\cite{jiang2013survey}, \eg Apriori based algorithms (\cite{APRIORI})
\end{itemize}

These graph mining tasks, which have relations among entities as the cornerstone of their design, are applicable to domains such as life sciences (\cite{STOCHPROT,GUI2,PROTPROT}), sociology and social networks (\cite{AA,LPSP,TOYO2}), collaboration analysis (\cite{DLP,LPCN,TIMEAWARELP}), business and product recommendation (\cite{RECOMM,LPCOLFIL}), and even law enforcement and anti-terrorism (\cite{SUPERVISEDTERROR,CLA,TERROR}).

The increased dimensionality of graph data sets often comes hand in hand with an increase in size. Together, large dimensionality and size, define the increasingly frequent family of domains known as large scale networks (\eg Twitter, the brain connectome or web graphs). Regardless of the underlying domain, computing large scale networks represents a challenge in terms of efficiency, parallelism and scalability. Efficiency, because computing models and hardware architectures are not optimized for handling graph data. Parallelism, because the size of large networks makes serial approaches unfeasible. And scalability because limited computational resources constrain the applicability of exhaustive, model-based solutions. Beyond the challenges on \emph{how} is the process implemented, the particularities of large scale networks also generate novel challenges on \emph{what} must be done with the data from a data mining perspective. A prime example of that is deciding how to evaluate the performance of the mining algorithms in this novel setting.

In this paper we focus on the challenge of faithfully evaluating a graph mining task, Link Prediction (LP), when working with large scale graphs. The goal of LP is to find new or missing edges within a given graph. By using LP one can directly grow any data set represented as a graph using the same graph language (\ie adding edges among vertices by using only previous edges and vertices). As a result one could apply LP algorithms to virtually any domain that can be represented as a graph without supervision. The complexity of achieving good performance on the LP task increases with the graph size, as does the problems at faithfully evaluating performance. When a graph grows linearly in vertices, the number of possible links within the graph grows quadratically. This defines a \emph{needle in a haystack} context where relevant or useful predictions are but a tiny fraction of all predictions. Keeping a good precision in this type of problem turns out to be very difficult, as the smallest false positive acceptance rate will amount to a huge absolute number of wrongfully predicted edges (\ie false positives). But in parallel, estimating the quality and applicability of results also becomes particularly difficult. 

In \S\ref{sec:context} we explore the current solutions provided by the data mining community, particularly in the context of test set construction and class imbalance. We explore the features of those methods for the particular case of LP in \S\ref{sec:evaluating}, and argue on the utility of popular approaches like ten-fold cross validation and precision-recall curves. Then in \S\ref{sec:CAUPR} we propose an adapted evaluation methodology, and show an empirical analysis on its impact when applied to several graphs. Conclusions are presented in \S\ref{sec:concl}.

This paper is an extended version of \cite{CCIA15}, improving the definition of the proposed evaluation methodology, analysing its properties in more depth, and adding an empirical comparison between the proposed methodology and current solutions (\S\ref{sec:CAUPR}). Further images and tables are provided to illustrate on the relevance of the contribution.

% The challenges explored in this paper refer to the LP problem on large graphs, but can be generalized to other large scale graph mining tasks. 

% The low precision of current LP results makes them hardly applicable to real contexts. As a result, the evaluation methodology found in the current state-of-the-art focuses on comparing the performance of algorithms, instead of comparing the applicability of their results. In this paper we discuss this issue...

\section{Evaluation context}\label{sec:context}

LP and the rest of graph mining tasks represent a new family of data mining algorithms. The particularities of these algorithms originate from the special nature of networks. Particularities that include data dimensionality, variable dependency, and often log-scale distribution of information. Even with these differences, one can find analogies between graph mining problems and general data mining problems. From a traditional data mining perspective, LP can be reduced to a binary classification problem between two classes: the positive class of edges that do or should exist, and the negative class of edges that do not and should not exist. Given a directed graph $G=(N,E)$, and all the possible edges in the graph (of size $|N*(N-1)|$), the problem of LP would be that of distinguishing between those edges that exist, $e\in E$, and those that do not, $e\notin E$. The analogy between LP and binary classification is accurate in most cases, as the target of LP is often to identify the positive class. Which in terms of graph mining is equivalent to finding and proposing missing links. For the remaining of this paper we will assume this mainstream case. 

In the bibliography there are a many methodologies available for the evaluation of a binary classification problem. These methodologies are typically discriminated based on the problem characteristics, which in the case of the LP classification problem are dominated by class imbalance.

\subsection{Test Sets}\label{sec:test_sets}

To evaluate a binary classifier empirically we require a test set. Given a graph, LP algorithms can propose a number of edges to be added to it, however, to validate the quality of those proposals, we need a set of edges known to be correct and missing from the graph. In evaluation, each predicted edge found in the test set is considered as a correct prediction, while each predicted edge not found is considered as a mistake. From these results one can then obtain performance indicators like \emph{precision} and \emph{recall}.

The main problem with tests sets is how to obtain them. In the case of LP, the best test set one can use is that which represents a natural extension of the graph being computed. This is feasible on temporally grounded domains. For example, for a graph composed from Wikipedia articles and the hyperlinks among them from 2012, we can obtain a natural test set by considering the links added to Wikipedia after 2012 (\cite{PHD}). Unfortunately, the domains and graphs having such incremental nature are rare. Instead, in most cases one must settle for the more drastic approach of randomly removing a number of edges from the graph in order to use them as test set. This yields other problems such as how to define a representative test set for the LP problem \cite{zhu2012uncovering}.

A frequent concern when one must split a set of data to produce a test set is representativeness. Typically, a random split cannot guarantee a prototypical distribution. The most frequent solution for avoiding bias is ten fold cross validation (10-fold CV). Within the LP problem, splitting data to build a test set will be often necessary (\cite{EVALLP}). However, as we show in \S\ref{sec:10fcv}, performing 10-fold CV is redundant. Hence, for all the test performed in this article we will use a random split of 10\% of edges on each graph to build the test set.

\subsection{Class imbalance}\label{sec:imbalance}

A recurrent type of real world graphs are scale-free networks, from protein interactions, to social networks or the WWW \cite{barabasi2009scale}, a type of network where degree distribution follows a power-law. This distribution implies a significant sparsity in the graph \cite{del2011all}, which becomes more severe as the graph grows. It is indeed hard to find real world networks where the average vertex degree is over fifty (\cite{DLP,NEWPERS,LP}), a feature consistent even as graphs grow to billions of vertices (\cite{TOYO1}). In the context of reducing the LP problem to a binary classification problem, scale-free networks results in a severe class imbalance, as the negative class becomes much larger in comparison to the positive class. As is well known, class imbalance can be a severely complicating factor in classification problems (\cite{SMOTE,UNDER,FEAT,RARITY}). 

The degree of class imbalance found on large graphs when performing LP is hard to overestimate, and even for non-scale-free networks, large, dense graphs are very hard to come by. To illustrate on the type of class imbalance medium and large graphs may have, Table \ref{tab:avg_epn} shows the topological properties of some real world graphs obtained from WordNet (\cite{KNOW}), the Cyc project (\cite{CYC}), the movie-related IMDb knowledge base (\cite{PHD}), and several web graphs from the Notre-Dame University (\cite{WEBND}), Stanford/Berkley universities (\cite{WEBSB}), a Google challenge (\cite{PHD}), and the Hudong, Baidu (\cite{KONECT}) and Wikipedia encyclopedias (\cite{DBP}). Notice how, in the \emph{best} case scenario, the class ratio is of 1 positive instance for every 11,382 negative instances. 

\begin{table*}[htb]
 \centering
\caption{Sample of average number of edges per vertex and class imbalance on real graphs}
\label{tab:avg_epn}
  \begin{tabular}{ c  r  c  l }
    \hline
    
\multicolumn{1}{l}{\bf{Data}} & \multicolumn{1}{l}{\bf{Number of}} & \bf{Average edges} & \bf{positive:negative}\\ 
\multicolumn{1}{l}{\bf{source}} & \multicolumn{1}{l}{\bf{vertices}} &  \bf{per vertex}   &   \bf{class ratio}  \\ \hline

%     \bf{Data source} & \bf{Number of vertices} & \bf{Average edges per vertex} & \bf{positive:negative class ratio}\\ \hline
      WordNet 	& 89,178 	& 15.66 	& 1:11,382 \\ \hline
      Cyc 	& 116,835 	& 5.9 	& 1:39,496 \\ \hline
      webND 	& 325,729	& 9.18 	& 1:70,867\\ \hline
      webSB 	& 685,230	& 22.18	& 1:61,775\\ \hline
      webGL 	& 875,713	& 11.64	& 1:150,217\\ \hline
      hudong 	& 1,984,484 	& 14.98	& 1:264,848\\ \hline
      baidu 	& 2,141,300 	& 16.72	& 1:257,667\\ \hline
      IMDb 	& 2,930,634 	& 5.12 	& 1:1,140,835 \\ \hline
      DBpedia 	& 17,170,894 	& 19.44	& 1:2,151,672 \\ \hline
  \end{tabular}
\end{table*}

The impact of class imbalance on classifiers was explored in \cite{IMBALANCE}, and authors concluded that this impact was largely reduced when all classes were of reasonable size. \emph{A priori} this should be good news for LP on large graphs, as its classes seem to be of \emph{reasonable} size; the positive class of all graphs shown in Table \ref{tab:avg_epn} is over 10,000 entities. Unfortunately, this assumption does not apply to the LP problem (\cite{NEWPERS}), and the reason for this is twofold. On one hand the imbalance found in LP on large graphs is several orders of magnitude larger than any imbalance tested in \cite{IMBALANCE}. Thus its impact may become significant at some point. On the other hand, LP is not a standard data mining classification problem, and given the small amount of information provided by each edge (\eg positive instances have no attributes), a class composed 30,000 elements could still be considered to be small. In reality, class imbalances of 1:10,000 or larger translate as a strong tendency towards false positive classification mistakes, as incorrectly accepting negative instances becomes almost inevitable. The main challenge of LP is therefore precision, a notion that should be taken into account by the evaluating methodologies.

\subsection{Evaluation under class imbalance}\label{sec:eval_under_ci}

Class imbalance is key in classification problems as it implies difficulties in predicting the small class. A small class that is in most cases the main target of the predictive process. Consequently there is a large and growing state-of-the-art on how to deal with class imbalance. A frequent approach of supervised or semi-supervised learning methods to overcome class imbalance is to equilibrate the training set through over-sampling, under-sampling or feature selection (\cite{SMOTE,UNDER,FEAT,RARITY}). Unsupervised LP algorithms cannot benefit from these solutions as adding or removing edges from the data set would equal to perform arbitrary classification, and there are no features to be removed beyond the existence of edges among vertices. As a result, for the LP problem one must focus only on those aspects of class imbalance that are relevant for unsupervised methods: deciding which metrics to use when evaluating and comparing the performance of binary classifiers for data sets with a large class imbalance. 

The most frequently used methods for classifier evaluation are based on accuracy. However, these methods are biased towards the classification of instances within the large class, making them inappropriate for imbalanced data sets (\cite{SMOTE,IMBALANCED,UNDER,FEAT}). Using them for LP would be almost analogous to measuring the capability of algorithms at predicting which edges should not be added to the graph, which is not the goal of LP. For data sets with large class imbalance, the most frequently used methodology is the Receiver Operating Characteristic (ROC) curve and the derived Area Under the Curve (AUC) measure (\cite{ROC}). The ROC curve sets the True Positive Rate (TPR) against the False Positive Rate (FPR), making this metric unbiased towards entities of any class regardless of their size. The AUC measures the area below the curve in order to compare the overall predictive performance of two different curves.

ROC curves are unbiased in imbalanced contexts, but their consideration of miss-classifications can result in mistakenly optimistic interpretations (\cite{ROC_PR,EVALLP}). When the negative class is very large, showing mistakes as relative to the negative class size (\ie FPR) can hide their actual magnitude, and make it complicated to assess the overall performance quality. From a practical perspective, most of the ROC curve is irrelevant when dealing with large class imbalance, as it represents completely unacceptable precisions. For example, one may consider that a classifier achieving a TPR of 0.95 (finding 95\% of all positive edges) and a FPR of 0.01 (incorrectly accepting 1\% of all negative edges) in the ROC curve demonstrates an excellent performance. However, for a data set with a positive:negative rate of 1:100 those results imply that the classifier accepts more negative edges than positive edges (\ie it has a precision smaller than 0.5). For domains with a 1:11,000 or worse ratio, like the ones shown in Table \ref{tab:avg_epn}, the limitations of the ROC curve become even more striking. In those even a FPR of 0.0001 implies a very poor precision/performance regardless of the TPR achieved.

\begin{figure}[!h]
\centering\includegraphics[width=3.5in]{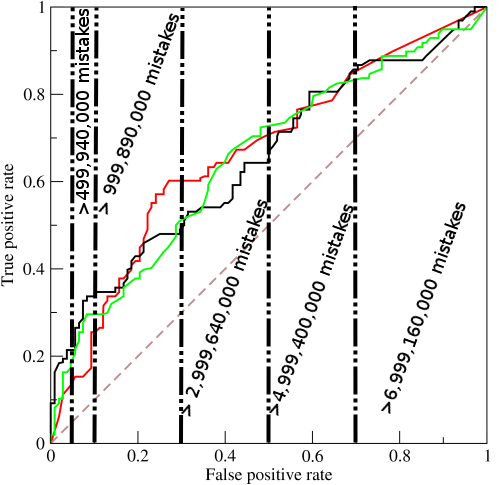}
   \caption{Example on the impact of imbalance on a ROC curve, showing the number of classification errors done at each FPR for a graph defined by N=100,000 and E=1,000,000. The true positive class size for this problem is 100,000.}
  \label{fig:roc1}
\end{figure}

Consider a theoretical graph defined by N=100,000 and E=1,000,000, for which we build a test set using 10\% of the available edges. The positive class size of this graph will be 100,000, the negative class size 9,998.9 million, and the imbalance ratio 1:99,989. Notice this graph is not particularly imbalanced (see Table \ref{tab:avg_epn} for comparison with graphs coming from real-world domains). A ROC curve for a LP algorithm on this theoretical graph could look like the one shown in Figure\ref{fig:roc1}. A FPR of 0.1 for such a graph (incorrectly accepting 10\% of negative class instances) would imply the wrong prediction of 999,890,000 edges, while our graph originally had 1,000,000 edges (900,000 after the test set split). Even with a FPR of 0.0001 a classifier would be making more false predictions than edges found in the graph. This simple example shows how most of the ROC curve is virtually useless for domains with a very large imbalance, which leads us to seriously question the utility of the associated AUROC measure in this context.

Precision-recall (PR) curves are an alternative to ROC curves. A PR curve is resistant to class imbalances as it focuses only on the performance achieved for the positive class (typically the small one), and does not show the number of correct classifications for the negative class. PR curves plot precision (\emph{y} axis) against recall (\emph{x} axis), directly showing the precision of the classifier. Precision can also be obtained from ROC curves, but is not straightforwardly visible. ROC and PR curves are related; a curve dominates another (it is above it) in the ROC space if and only if it also dominates it in PR space, but are not equivalent; a classifier that optimizes on ROC space will not necessarily also optimize on PR space (\cite{ROC_PR}). One particularly relevant difference between ROC and PR curves regarding the interpretation of predictive performance is on how errors are represented. While ROC curves show miss-classifications as relative to the total number of negative cases, PR curves show miss-classifications as relative to the total number of predictions done.

An illustration on the impact of these differences is the curves defined by a random classifier, which always performs poorly in an imbalanced data set. The ROC curve always represents the random classifier as a straight line between points $(0,0)$ and $(1,1)$, regardless of class imbalance, with all better than random classifiers represented as lines above that diagonal. PR curves on the other hand represent random classifiers in imbalanced data sets a flat line on the \emph{x} axis, as their precision in imbalanced settings is always close to zero. This alone shows that PR curves can provide richer characterizations of classifiers for imbalanced data sets.

\section{Evaluating link prediction}\label{sec:evaluating}

Current solutions for performance evaluation, like the ones shown in \S\ref{sec:context}, have severe limitations when applied to large graph mining problems. Issues like test set representativeness, or the evaluation under class imbalance, reach a new degree of relevance when considering problems like LP on large networks. In this section we discuss these problems in depth and propose solutions fitting our LP problem.

\subsection{Representativeness of test sets}\label{sec:representativity}

The use of a test sets to evaluate LP implies the assumption that the test set (the prediction of which is evaluated by the curves) faithfully represents the \emph{correct} edges missing from the graph. Or in other words, that all edges not found in neither the graph nor in the test set, are wrong. In certain cases, where the graph topology is stable, this may be an accurate assessment. For example, a graph obtained from WordNet data (as shown in \cite{KNOW}) can be considered as almost perfect, since WordNet relations have been identified, discussed and implemented by linguists for decades. In other cases though test sets are an imperfect measure of the right edges missing from the graph. Consider for example a graph obtained from Wikipedia articles and hyperlinks, in which the pagelinks among Wikipedia articles from 2012 are used as training and the new pagelinks added on 2013 are used as test. This graph is clearly incomplete, as new links are being added every day. The Wikipedia grows continuously and the fact that a link is not implemented so far does not mean it is wrong. As a result, one must take into account that some of the edges predicted, not found in the test set and labelled as mistakes, will in fact be correct predictions corresponding to edges not yet added to the graph.

Using a test set which does not fully represent the target class implies an underestimation of performance, as the predictions being made outside of the test set will always (and not always correctly) be considered as mistakes. Nevertheless, since this limitation applies to all the methods being evaluated (assuming all methods are evaluated using the same test set), it can be argued that the resultant performance indicators remain valid for comparative purposes. That is, we can still find out which LP algorithm works better. The unavoidable shortcoming of representativeness comes when evaluating the precision of a score in the context of applicability. That is, we cannot be sure of how well performs the best LP algorithm. The only way to obtain a faithful, non-comparative evaluation of performance of a single LP algorithm would be a hand-made validation. One could achieve an approximate solution by performing a sampling process of all edges predicted, manually evaluating the sampled edges as correct or incorrect predictions, and then extrapolating the performance obtained on the sample to the rest of the graph. There are several aspects to keep in mind with this solution. First of all, the sampling needs to be large for the extrapolation to be faithful, which equals to many hours of manual labelling. And second, the sampling would have to be done at several thresholds so that extrapolations are representative of the whole curve. Sampling may therefore be the only accurate evaluation methodology for estimating predictive performance of a given score on a specific domain, at the price of a huge amount of manual labelling hours.

\subsection{10-fold CV}\label{sec:10fcv}

10-fold CV is a commonly used technique for reducing variance in test set construction and improving representativeness. Although 10-fold CV is almost universally expected when using test sets that are a random portion of a complete data set, we argue that it is not needed when performing large graph mining. The main reason behind that argument being the large size of these domains, which naturally avoid variance. To assess the utility of 10-fold CV we test a webgraph obtained from a Google challenge, composed by 875,713 vertices and 5,105,039 edges. This particular graph could be considered to be medium sized, as it is easy to find much larger ones (see Table \ref{tab:avg_epn}). The conclusions obtained for this graph could be extended, even with more reliability, to larger graphs. 

\begin{figure*}[thb]
\centering
  \includegraphics[width=0.9\linewidth]{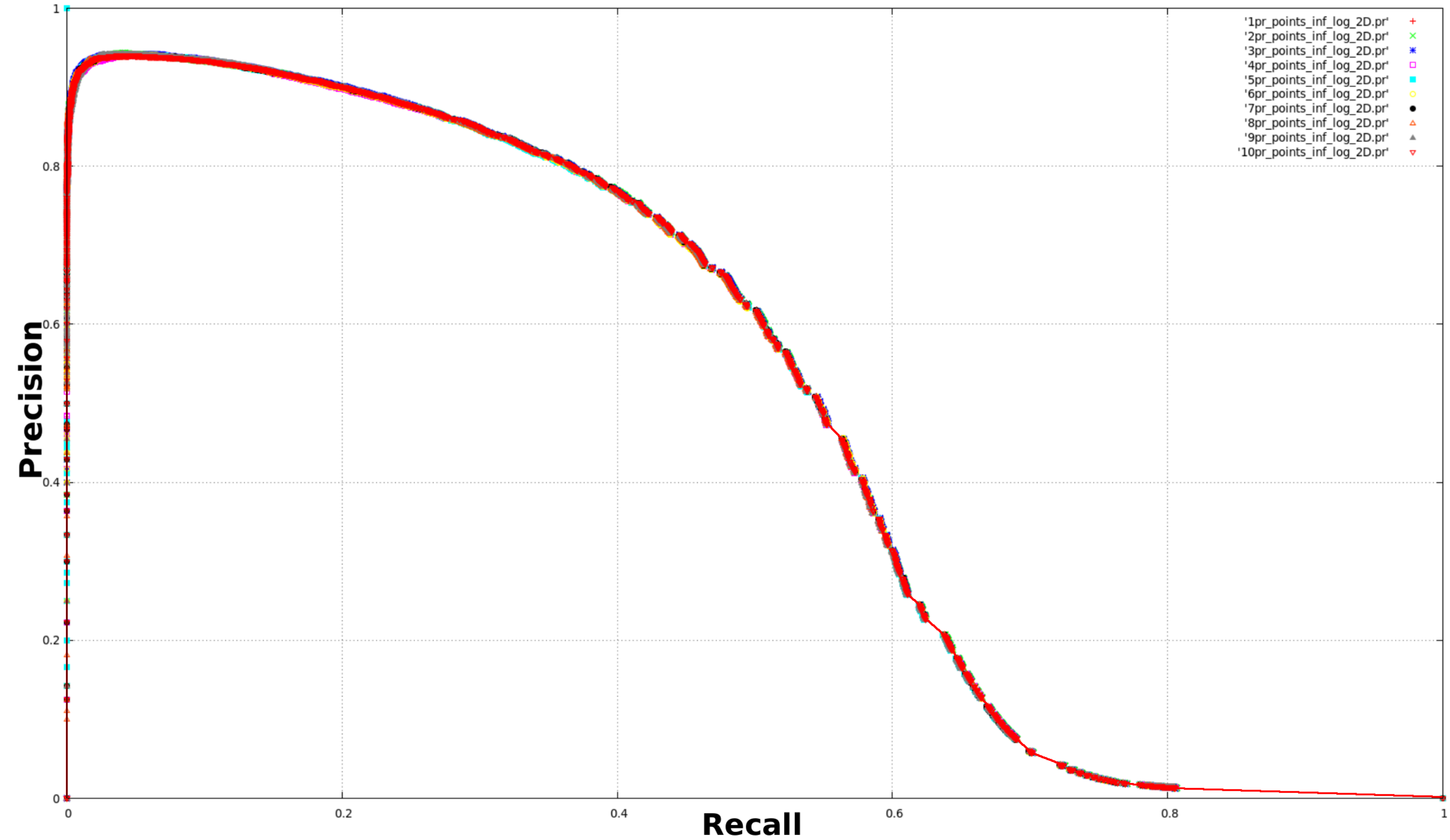}
   \caption{Ten precision-recall curves for the INF\_LOG\_2D LP algorithm when applied to 10 random splits of the Google challenge webgraph. The ten curves are clearly overlapped, showing minimal variance among random splits.}
  \label{fig:10test}
\end{figure*}%

A random 10\% test set of the Google challenge webgraph is composed by 510,503 edges. We obtain ten different random test sets from this graph, and use each one of them to evaluate seven different LP algorithms. As a result we obtain ten PR curves (as these are preferred to ROC curves, see \S\ref{sec:eval_under_ci}), for seven different LP algorithms. Algorithms used are RA (Resource Allocation), AA (Adamic-Adar), CN (Common Neighbours), INF, INF\_LOG, INF\_2D and INF\_LOG\_2D, all of which are described in \cite{PHD}. In Figure \ref{fig:10test} we show the ten curves belonging to one of those algorithms (INF\_LOG\_2D), illustrating the minimal variance found among curves. The ten curves are virtually identical, which implies that variance among random splits is irrelevant. To empirically validate this assertion, in Table \ref{tab:10test} we show the AUPR of the seventy curves obtained, ten for each algorithm. Results show that the variance of 10 executions using 10 randomly selected tests sets is very low. In fact, the standard deviation represents a 0.32\% of the mean value in the worst case (algorithm \#3). Such a low variance is the result of having a large test set, which, given the law of large numbers, will tend towards a stable sample. In this context it seems clear that performing 10-fold CV is not necessary, as a single run is a representative and accurate sample of performance.

\begin{table*}[tbh]
 \centering
\caption{Using the Google challenge webgraph, AUPR obtained by seven different algorithms on ten different and randomly split test sets. The minimum and maximum value among the ten splits, and the standard deviation.}
\label{tab:10test}
  \begin{tabular}{ l  l  l  l  c }
    \hline
    \bf{Algorithm} & \bf{Min. AUPR} & \bf{Max. AUPR} & \bf{Mean} & \bf{Std. Dev.}\\ \hline
      AA 	&0.0892558 	&0.0899991 	&0.08971357 	&0.0002287783\\ \hline
      CN 	&0.10017 	&0.10083 	&0.1005145 	&0.0001902058\\ \hline
      RA 	&0.0618143 	&0.0625483 	&0.06225763 	&0.0002040158\\ \hline
      INF 	&0.128201 	&0.128857 	&0.1285072 	&0.0001879318\\ \hline
      INF\_LOG 	&0.124934 	&0.125385 	&0.1251525 	&0.0001210622\\ \hline
      INF\_2D 	&0.419577 	&0.421239 	&0.4204078 	&0.0004921798\\ \hline
      INF\_LOG\_2D 	&0.491902 	&0.4935 	&0.4925985 	&0.0005283041\\ \hline
  \end{tabular}
\end{table*}

Regardless of these results, performing 10-fold CV is not a wrong or misguiding strategy. Our argument here is that 10-fold CV is not required in order to consider some results representative. This fact is particularly relevant due to the computational cost of computing large scale graphs. Building test sets and running graph mining algorithms on them is typically expensive in computational terms. Hence, the physical resources and time spent doing ten equivalent executions could be use more efficiently elsewhere.

\subsection{Precision-Recall curves in link prediction}\label{sec:eval_LP}
 
Most research on LP use ROC curves (\cite{CLA,LPCE,LP,LPCN,LPSP}) or PR curves (\cite{DLP,TENSOR}) for evaluation, but for the reasons discussed in \S\ref{sec:eval_under_ci} we find PR curves to be more appropriate. PR curve shows the performance of a classifier at various thresholds: at the left part of the curve are the high-certainty predictions where precision is higher, while at the right part of the curve are low-certainty predictions where recall grows at the expense of a lower precision. Through the PR curve one can see which classifier performs better at each threshold. The derived AUPR metric of the PR curve on the other hand determines which classifier performs better overall, when all thresholds are considered at the same time with the same importance. Due to this last point, we find the AUPR score to be sub-optimal for evaluating the applicability of results. Given the imbalance of the graphs used (see Table \ref{tab:avg_epn}), a large part of the PR curve represents very low precisions. As recall grows precision can quickly reach levels unacceptable from a practical point of view. At this point one must consider which results are worth taking into account for evaluating performance. If we intend to achieve an applicable methodology we should focus on its performance where it matters, when a \emph{reasonable} number of mistakes are being done. At extremely low precisions (\eg 0.01\%) results are likely to be useless, and therefore should not be taken into account (certainly not with equal weight) into the evaluation. For this reason in \S\ref{sec:CAUPR} we propose a modified version of the AUPR measure with the goal of focusing on applicability.

% one can linearly interpolate ROC, but not PR. In PR linear interpolation is a mistake that yields an overly-optimistic estimate of performance\cite{ROC_PR}. PR curves can be interpolated, but the process ain't that easy. its a curve. The effect of incorrect interpolation on the AUC-PR is especially pronounced when two points are far away. link this with the matter of numeric precision, show the number of points defining each curve for each number of decimals

\section{Constrained AUC}\label{sec:CAUPR}

The goal of classic binary classification is to fully discriminate two sets. This is idoneous when dealing with balanced or \emph{typically} imbalanced domains, but becomes problematic for domains with class imbalances in the order of millions (\eg see Table \ref{tab:avg_epn}). In this context, exhaustively discriminating the two classes becomes exceedingly complicated, which eventually renders a portion of the results obtained useless (see Figure\ref{fig:roc1}). Evaluation methodologies are unaware of the actual utility of each portion of the results, and combine the evidence provided by all results equally. Consequently, as the portion of results that is useless grows, the usefulness of the evaluation methodologies results decreases.

The LP problem is one where this disjunction between usefulness and classification performance takes place. As a solution we argue that the goal of LP is to produce high certainty and high utility predictions, instead of fully discriminating two sets. Indeed, LP does not need to classify most edges correctly in order to be useful, while trying to correctly classify all possible edges is a virtually impossible task due to the complexity of the problem. Instead, for the sake of making it useful for real world applications, LP should try to identify as many positive edges as possible, while keeping the number of false positives within an acceptable range.

% Evaluating LP performance in the context of applicability, so that we can determine which scores produce the best results, is . The classic AUC measure equally considers the performance of a classifier at any threshold, therefore evaluating which classifier performs the best overall. However, we consider that the performance of LP scores at low thresholds, where \emph{billions} of mistakes are done, is irrelevant for assessment purposes. 

To formally evaluate performance in terms of \emph{usefulness} we first need to ground that subjective term. Since every domain, application and even user may have its own definition of it, we decide to define instead a minimal threshold which guarantees that all useful results are beyond it. If the hypothesis is accepted, all relevant results will be provided and accounted for, and utility, although not optimized, will be improved. The hypothesis we propose is as follows:

\begin{hypothesis}\label{ass}
 Given a link prediction process applied on a graph $G=(V,E)$, once the number of false positives is equal or larger than $|E|$, all further predictions become irrelevant.
\end{hypothesis}

The idea behind this hypothesis is that, given a data set $X$, we will rarely accept any result which includes a number of mistakes as large as $X$ itself. This is a conservative approach that may hold for most applications and domains.

Based on this hypothesis we propose the Constrained AUC score (CAUPR when applied to the PR curve) with the goal of evaluating LP scores based only on the predictions produced while keeping an acceptable number of mistakes (\ie less than the graph size). The CAUPR is analogous to the traditional AUPR measure, computing the AUC of the PR curve where the number of non-existing edges mistakenly accepted by the score (\ie false positives) is equal or lower than the total number of edges in the graph. Once the number of false positives is larger than the number of edges, the CAUPR for the rest of the curve equals 0. In practice, the CAUPR is a subset of the AUPR, starting from the high confidence predictions (left side of the PR curve) and ending when a given threshold is reached. 
% Formally,\\
% 
% 
% \begin{definition}
%  A link prediction score LP produces a precision-recall curve PR when applied to a graph G=(V,E). Considering that the negative class of this problem is N, we define a minimal utility threshold $\tau$ such that $\tau=\frac{|E|}{|N|}$.\\
%  
%  The CAUPR of LP on G is thus $CAUPR\subseteq AUC_[]$
% \end{definition}

\begin{figure*}[htb]
  \centering
  \includegraphics[width=1\linewidth]{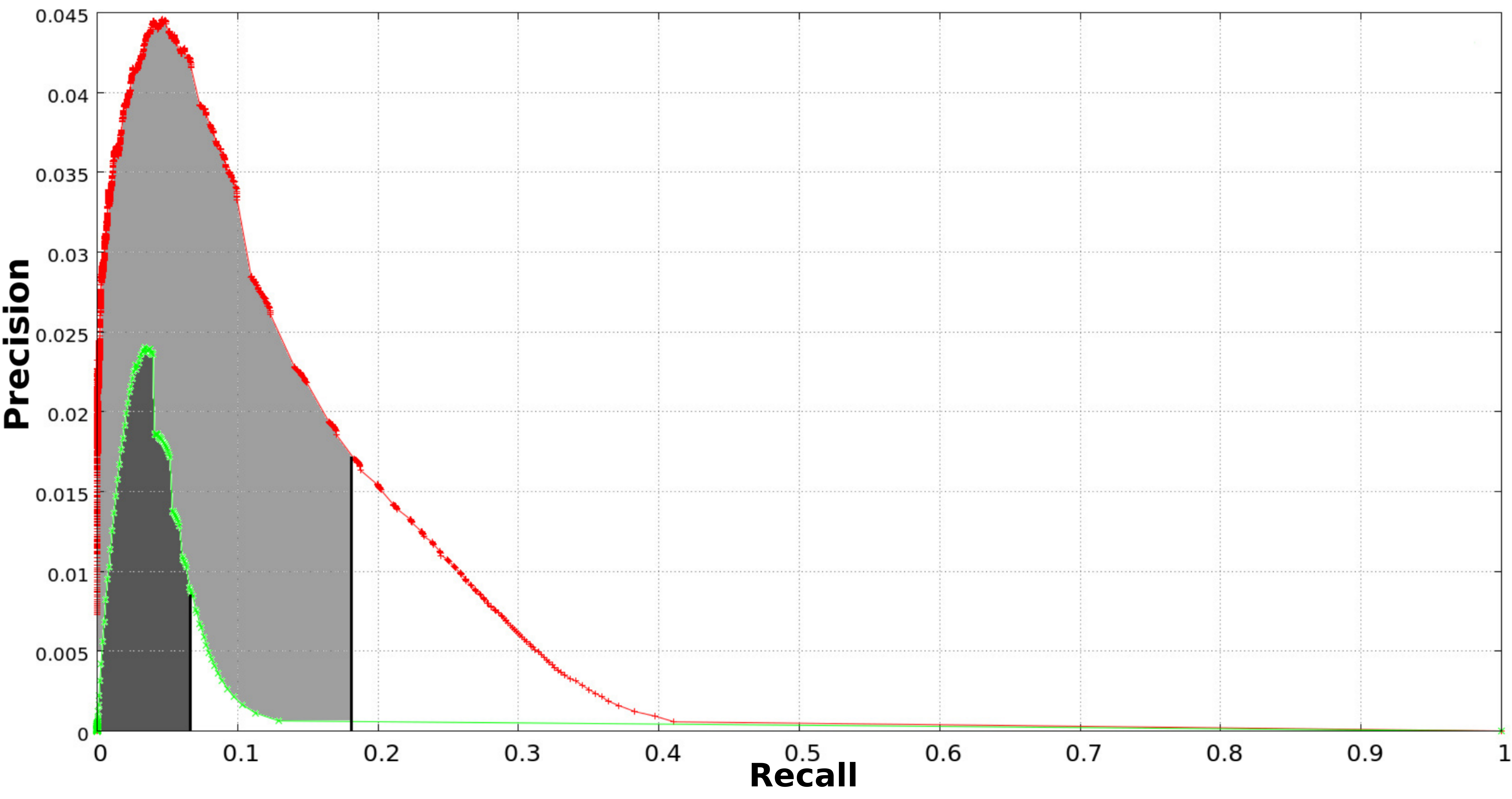}
   \caption{Precision-recall curve of the RA LP score on the Cyc graph (upper curve), and precision-recall curve of the same algorithm on the IMDb webgraph (lower curve). Grey area shows the respective CAUPR. The CAUPR threshold for the Cyc graph corresponds to a higher recall value (approximately 0.18) due to the higher precision obtained in this graph, despite Cyc being a smaller graph than IMDb. The CAUPR threshold for the IMDb graph corresponds to a recall value of approximately 0.06, thus by the time it finds 6\% of all positive edges, the algorithm has made as many mistakes as edges in the IMDb graph.}
  \label{fig:CAUPR_example}
\end{figure*}

As an example, see Figure \ref{fig:CAUPR_example}, where the PR curves of a LP score are shown for two different graphs. The vertical cut on each curve represents the location of the CAUPR threshold for each particular data set and score, limiting the CAUPR to the area at the left of the threshold (coloured in grey), whereas the AUPR considers the whole curve.

% formula

\subsection{CAUPR properties}

The goal of the CAUPR performance measure is to avoid the evaluation of irrelevant parts of the PR curve. For that purpose CAUPR defines a threshold $x$ at which a given algorithm is accepting more false positives than edges available in the graph. For PR curves, $x$ indicates the maximum recall an algorithm can obtain before reaching the threshold, and splits the PR curve in two. The PR sub-curve in the recall interval $[0,x]$ will be the one CAUPR will take into account, while the PR sub-curve in the recall interval $[x,1]$ will be the difference between the CAUPR and the AUPR. Significantly, $x$ is in the interval $[0,1]$. It can be zero, if the first $|E|$ predictions made by the LP algorithm are mistakes, but it can also be one, if all true positives are found before $|E|$ false positives are accepted. In this last case the CAUPR ignores nothing of the curve, and is equal to the AUPR. Consequently, the CAUPR and AUPR measures will only differ when the LP algorithms do not perform well enough (as defined by Hypothesis \ref{ass})

\begin{figure}[bt]
  \centering
  \includegraphics[width=0.9\linewidth]{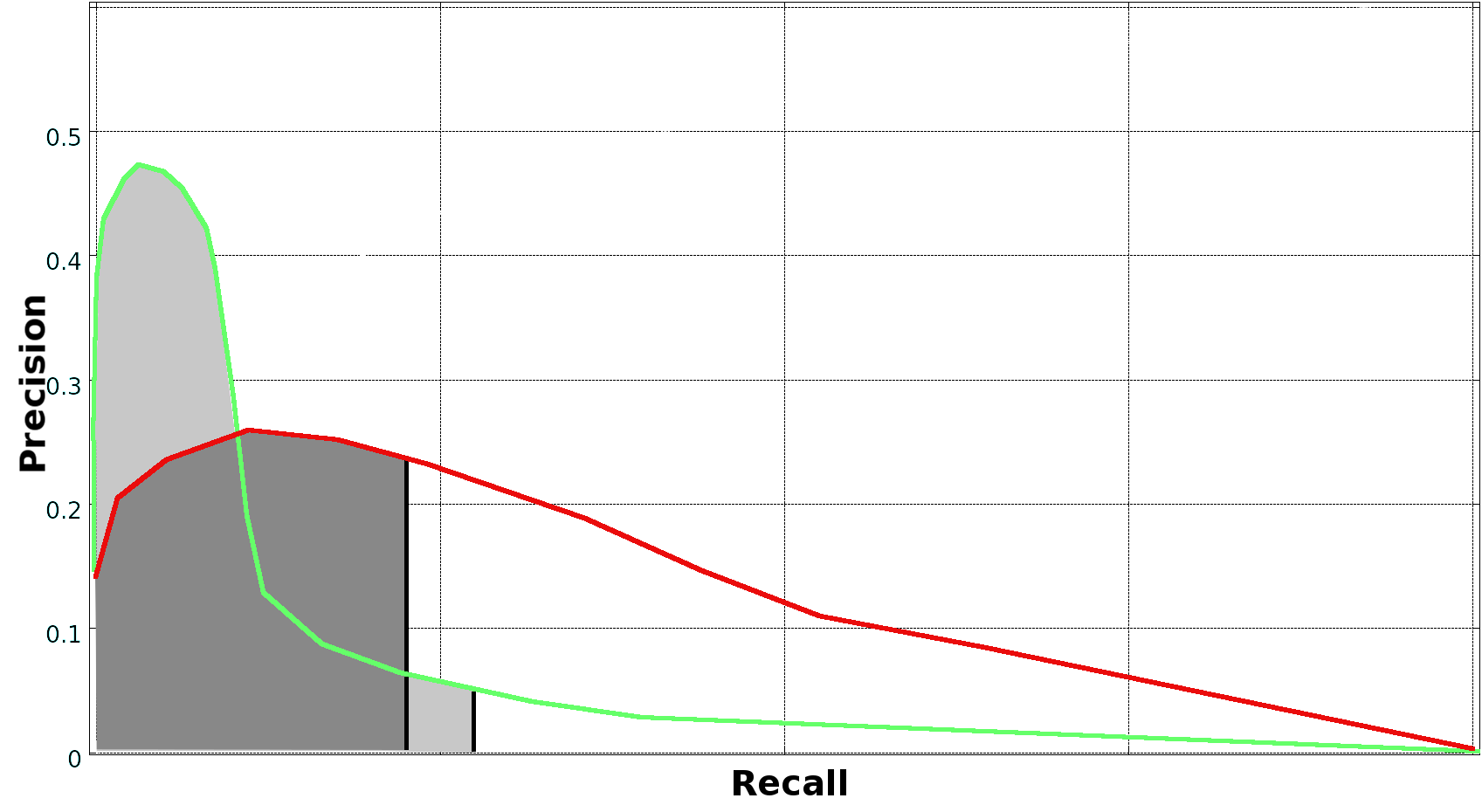}
   \caption{Illustration of how AUPR and CAUPR can produce contradictory comparisons due to different precision-recall curve shapes. Grey area represents CAUPR. In this example one curve has the larger AUPR, while the other has a larger CAUPR.}
  \label{fig:CAUPR_profile}
\end{figure}

The LP scores that may be penalized by CAUPR in comparison with the AUPR are those which outperform their competitors on the \emph{irrelevant} part of the curve. Since that area, if existent, is located at the right side of the curve, and since the PR curve is monotonically decreasing, the CAUPR will penalize the scores producing more horizontal PR curves. On the other hand, LP scores which make more accurate predictions at the beginning of the curve, when the number of false positives is still assumable, but which quickly lose precision (\ie those with a more vertical PR curve), will be the ones to benefit from the CAUPR. Consider the PR curves of Figure \ref{fig:CAUPR_profile}. The more horizontal curve (H) has a larger AUPR than the more vertical curve (V), but V outperforms H in CAUPR. This is due to the higher precision obtained by V at high confidence predictions, which causes a larger portion of V's AUPR to be considered by CAUPR. H, on the other hand, outperforms V at higher recall values, and mostly when the number of false positives is already larger than the graph itself, beyond the CAUPR threshold.

\begin{table}[bt]
\centering
\caption{For 9 graphs, 2nd to 4th columns show best AUPR obtained by one of CN, AA and RA algorithms (\ie algorithm of reference), and the CAUPR obtained by that same algorithm. Remaining columns show the AUPR and CAUPR obtained by the other two algorithms as a percentage of the results obtained by the algorithm of reference. \textit{Impact} column shows the difference of the percentages, summarizing how CAUPR alters the comparison between the algorithm of reference and the other two LP algorithms.}
\label{CAUPR-auc-tab}
\begin{tabular}{c|cllclrr}
\hline

Graph & Ref.	& AUPR    	& CAUPR       	& Alg.	& AUPR as 	& CAUPR as 	& \textbf{Impact}\\
      & Alg. 	& of ref.	& of ref.	& 	& \% of ref.	& \% of ref.	&\\  \hline

% \multicolumn{1}{l|}{Graph} & \multicolumn{1}{l}{Ref.} & AUC    & CAUPR       & \multicolumn{1}{c}{Alg.} & \multicolumn{1}{c}{AUC as} & \multicolumn{1}{c}{CAUPR as} & \multicolumn{1}{c}{\textbf{Impact}}\\ 

% \multicolumn{1}{l|}{} & \multicolumn{1}{l}{Alg.} &  of ref.   &   of ref.     & \multicolumn{1}{c}{} & \multicolumn{1}{c}{\% of ref.} & \multicolumn{1}{c}{\% of ref.} & \multicolumn{1}{c}{\textbf{}} \\ \hline
WN         & RA  & 0,0487 & 0,0383 & AA  & 46,3\%  & 40,7\%  & \textbf{-5,6\%}   \\
	   &     &        &        & CN & 13,1\%  & 12,7\% & \textbf{-0,4\%}                                 \\ \hline

Cyc        & RA  & 0,0076 & 0,0058 & AA  & 79,5\%  & 83,5\%  & \textbf{+4,0\%}   \\
&     &        &        & CN & 19,0\%  & 23,2\% & \textbf{+4,2\%}                                  \\ \hline

WebND      & CN  & 0,3185 & 0,3158 & RA  & 66,4\%  & 65,8\%  & \textbf{-0,6\%}   \\ 
&     &        &        & AA & 99,4\%  & 99,0\% & \textbf{-0,4\%}                                 \\ \hline

WebSB      & RA  & 0,0549 & 0,0460 & AA  & 40,3\%  & 40,8\%  & \textbf{+0,5\%}   \\ 
&     &        &        & CN & 30,3\%  & 32,7\% & \textbf{+2,4\%}                                  \\ \hline

WebGL      & RA  & 0,1003 & 0,0921 & AA  & 89,2\%  & 88,6\%  & \textbf{-0,6\%}   \\ 
&     &        &        & CN & 62,0\%  & 60,8\% & \textbf{-1,2\%}                                 \\ \hline

IMDb       & RA  & 0,0015 & 0,0011 & AA  & 62,7\%  & 78,2\%  & \textbf{+15,5\%}  \\ 
&     &        &        & CN & 43,2\%  & 54,8\% & \textbf{+11,6\%}                                 \\ \hline

Hudong     & CN  & 0,0074 & 0,0072 & RA  & 30,0\%  & 21,8\%  & \textbf{-8,2\%}   \\ 
&     &        &        & AA & 74,6\%  & 69,7\% & \textbf{-4,9\%}                                 \\ \hline

Baidu      & RA  & 0,0031 & 0,0016 & AA  & 92,5\%  & 94,6\%  & \textbf{+2,1\%}   \\ 
&     &        &        & CN & 57,1\%  & 66,6\% & \textbf{+9,5\%}                                  \\ \hline

DBp    & AA  & 0,0005 & 0,0003 & RA  & 28,5\%  & 351,8\% & \textbf{+323,3\%} \\ 
&     &        &        & CN & 67,8\%  & 73,8\% & \textbf{+6,0\%}                                    \\ \hline
\end{tabular}

\end{table}

Another relevant feature provided by Hypothesis \ref{ass} is domain adaptation. By considering the number of edges in the graph as threshold, the CAUPR is affected by graph properties such as density and size (\ie large graphs will accept more mistakes than small ones, dense graphs will accept more mistakes than sparse ones). This is an interesting feature not found in the AUPR measure: AUPR evaluates the predictions done on a graph with $N$ vertices and 1,000 edges and the predictions done on a graph with $N$ vertices and 100,000 edges under the same conditions, as if these two problems were equally difficult. A clearly unrealistic assumption that complicates the interpretability of results. CAUPR, on the other hand, implicitly incorporates the size and density of the graph into the evaluation, allowing more concessions when they are acceptable by the domain, according to Hypothesis \ref{ass}). 

% Thus, CAUPR results across graph domains can be compared with a certain degree of neutrality.

\subsection{CAUPR impact}

To empirically evaluate the impact of the CAUPR measure we use three well known LP algorithms: Common Neighbours (CN, \cite{CN}), Adamic Adar (AA, \cite{AA}) and Resource Allocation (RA, \cite{RA}). For each of those algorithms we compute their AUPR and CAUPR when applied to the nine graphs described in Table \ref{tab:avg_epn}. The algorithm obtaining the best AUPR on each graph (\ie the algorithm of reference) and its corresponding AUPR and CAUPR values can be seen in the first four columns of Table \ref{CAUPR-auc-tab}. Each of the three LP algorithms obtains the best AUPR score on at least one of the nine graphs.

For each of the nine graphs we compare the results obtained by the algorithm of reference with the results obtained by the remaining two algorithms. We show the AUPR and CAUPR values of these two algorithms as a percentage of the values of reference, so that for example, for the WordNet graph, the AUPR of the AA algorithm is shown to represent a 46,3\% of the AUPR of the algorithm of reference (see 2nd row, 6th column of Table \ref{CAUPR-auc-tab}). By also showing the percentage relation for the CAUPR measure, we can see the relative difference between both measuring methods. Using the same example, since the CAUPR of the AA algorithm is 40,7\% of the CAUPR of RA, the difference is -5,6\% points. Thus, the difference in performance between the AA and RA algorithms is 5,6\% larger according to the CAUPR measure than for the AUPR measure.

The \textit{Impact} column of Table \ref{CAUPR-auc-tab} shows the changes in performance according to CAUPR, and shows how this measure may provide a relevant variation in the evaluation of LP for graphs with 100,000 vertices or more. Significantly, the variation provided by the CAUPR measure does not benefit any of the three algorithms: all of them have positive and negative differences. Results also indicate that variations tend to increase with the graph size, since larger graphs typically have larger imbalances, which often imply a lower recall threshold for the CAUPR measure. Clearly, having a lower recall threshold makes it easier (though not necessary) for the CAUPR and AUPR measures to differ. 

Table \ref{thresholds} shows the recall thresholds for the CAUPR measure of every graph and algorithm tested. This table gives a measure of the portion of the PR curve that is being disregarded by the CAUPR method. A threshold of 0.1 implies that 90\% of the curve is outside the CAUPR range, and therefore not considered in the CAUPR evaluation. The impact of the CAUPR measure, powered by the class imbalance, is highlighted by the fact that for five of the nine graphs a majority of the curve is irrelevant according to Hypothesis \ref{ass}.

One remarkable result to be considered is the variation on the largest graph used, the DBpedia graph. In this domain, the AA algorithm outperforms the rest according to the AUPR measure. However, according to the CAUPR measure the RA algorithm is best instead, with a three times larger CAUPR. To analyze these results lets first consider the DBpedia graph, which has the largest class imbalance of those graphs here considered, with more than 2 million negative edges for each positive one. As shown in Table \ref{thresholds}, RA performs very well on the DBpedia graph, reaching a recall of 26\% before the threshold of mistakes is attained. Comparably, AA retrieves only a 5\% of all positive edges by the time it reaches the threshold. Nevertheless, AA seems to outperforms RA for the portion of the curve beyond the threshold, thus obtaining a higher AUPR value. An example of PR curves with this kind of behaviour are illustrated in Figure \ref{fig:CAUPR_profile}, and are a showcase of the relevance of the proposed CAUPR measure.

\begin{table}[bt]
\centering
\caption{For 9 different graphs, CAUPR threshold showing at which recall value the number of mistakes is larger than the graph size. AUC beyond this recall value is not considered by the CAUPR measure.}
\label{thresholds}
\begin{tabular}{llll}
\hline
Graph   & \begin{tabular}[c]{@{}c@{}}RA CAUPR \\ Recall  \\ Threshold\end{tabular} & \begin{tabular}[c]{@{}c@{}}AA CAUPR\\ Recall \\ Threshold\end{tabular} & \begin{tabular}[c]{@{}c@{}}CN CAUPR\\ Recall \\ Threshold\end{tabular} \\ \hline
WN      & 0.539892                                                                & 0.276364                                                              & 0.147471                                                              \\ \hline
Cyc     & 0.183218                                                                & 0.144625                                                              & 0.092161                                                              \\ \hline
WebND   & 0.673279                                                                & 0.636334                                                              & 0.558378                                                              \\ \hline
WebSB   & 0.509361                                                                & 0.234861                                                              & 0.159888                                                              \\ \hline
WebGL   & 0.619519                                                                & 0.564522                                                              & 0.474023                                                              \\ \hline
IMDb    & 0.073010                                                                & 0.048817                                                              & 0.034015                                                              \\ \hline
Hudong  & 0.073233                                                                & 0.081013                                                              & 0.069519                                                              \\ \hline
Baidu   & 0.096464                                                                & 0.095531                                                              & 0.076136                                                              \\ \hline
DBpedia & 0.267792                                                                & 0.055546                                                              & 0.039784                                                              \\ \hline
\end{tabular}
\end{table}

\section{Related Work}

To the best of our knowledge, there are no previous proposals on how to adapt standard evaluation methods (\ie PR curves) to the particularities of large-scale LP. Similar solutions to our own (that of using a sub-part of the PR curve) have been previously considered for ROC curves on other contexts, particularly in the domain of diagnostic medicine, where several authors have considered the possibility of using only a partial ROC curve \cite{ma2013use,zhou2009statistical}. In this field, the metrics used to cut the ROC curve are often clinical relevance and clinical application. Rather differently, our proposal constrains the PR curve based on a domain agnostic measure: the input data set imbalance. The methodology we propose is thus applicable to virtually any LP evaluation problem, regardless of the data origin.

\section{Conclusions}\label{sec:concl}

% Graph mining algorithms, those that analyze and exploit inter-entity relations, provide distinct capabilities when compared to traditional data mining, due to the special nature of graph-like data (\eg high-dimensionality). The properties of graph data also imply particularities in most of the data processing workflow, such as scalability, parallelism and even performance evaluation. In here we discussed how to the LP problem of finding missing edges in a graph 

Two of the most disturbing features of real world graphs for the evaluation of LP algorithms are their size and scale-free topology. Medium sized graphs (\eg up to a few million vertices) are hard to compute through exhaustive algorithms, and have motivated the design of graph-specific parallel models (\eg \cite{PREGEL}). However, this same size can also simplify certain data mining steps, such as assessing test set construction as an independent dataset. This assessment, which is often implemented through 10-fold cross-validation, is actually avoidable in medium and larger graphs, since the size of a random 10\% split (\eg hundreds of thousands of vertices) already guarantees the construction of a stable sample (see Table \ref{tab:10test}). These results were consistent for graphs between 100,000 vertices and 17 million vertices, and can be extended to any graph larger than those. Avoiding cross validation can entail significant savings in terms of computational resources, allowing one to reduce the cost of every performance evaluation process by a factor of ten (assuming we were to use 10-fold cross-validation).

The second graph feature which is particularly relevant for evaluation is related with the scale-free topology of many real world graphs. Since LP can be reduced to a binary classification problem, where one tries to separate a positive class (the edges missing from the graph which should be added) from a negative class (the edges missing which should not be added), the scale-free topology implies a huge imbalance between both classes. In fact, imbalance reaches a degree rarely found in the bibliography, where, for every positive instance, there are tens of thousands or even millions of negative ones. Significantly, imbalance becomes larger as graphs do, making this an issue for current and future graph mining applications.

One side effect of huge class imbalance relates with the evaluation methodology being used. Binary classification problems are often evaluated through ROC curves, which plot TPR against FPR. FPR is however an uninformative performance scale in highly imbalanced domains, as most of the curve implies a huge amount of false positives (see illustrative Figure \ref{fig:roc1}). For LP in medium or large graphs, PR curves provide a much more realistic performance measure, since these curves plot precision against recall, directly displaying the number of false positives being done.

Unfortunately, using the PR curve does not guarantee the utility or correct interpretability of results, particularly if using the associated AUC measure. The large imbalance found in LP for large graphs often results in small precision values, which only get worse as recall grows. As a result, a significant part of the AUPR measure may correspond to predictions found at the cost of an unassumable amount of mistakes, and thus poorly represent applicable performance. To tackle this problem, we define a constrained version of the AUPR measure, CAUPR, by setting a conservative threshold for what number of mistakes are assumable. This threshold is based on the graph size (\ie we can accept as many false positives as edges in the graph), which provides several interesting properties. For example, the CAUPR may be equal to the AUPR, if performance is good enough, but it can also be zero if performance is very poor. Also, the CAUPR adapts to graph size and density, being more flexible when the domain allows. Nevertheless, the use of the CAUPR measure requires of the acceptance of Hypothesis \ref{ass}, which should be considered on a case by case basis.

Our empirical comparison between the AUPR and the CAUPR measures shows significant variances between both performance metrics, which tend to increase with graph size. On the evaluation of three LP algorithms, AUPR and CAUPR differ on which is the best one when applied to the largest graph computed (DBpedia, 17 million vertices), showcasing the relevance of the AUPR performance metric for LP evaluation on large and highly imbalanced graphs.

\section*{Acknowledgements}
This work is partially supported by the Joint Study Agreement no. W156463 under the IBM/BSC Deep Learning Center agreement, by the Spanish Government through Programa Severo Ochoa (SEV-2015-0493), by the Spanish Ministry of Science and Technology through TIN2015-65316-P project and by the Generalitat de Catalunya (contracts 2014-SGR-1051).

\printbibliography

\end{document}